\def\selectedoptions{final}

\documentclass[
   \selectedoptions
  ]
  {aipproc}

\layoutstyle{6x9}

\SetInternalRegister\hbadness{8000} 

\newcommand\doingARLO[2][]{%
  \ifx\mmref\undefined #1\else #2\fi
}

\begin{document}

\title{Lepton pair production in a charged quark gluon plasma}


\author{A. Majumder}{
  address={Department of Physics, McGill University, Montreal, QC,
Canada H3A 2T8},
  email={majumder@physics.mcgill.ca},
}

\author{C. Gale}{
  address={Department of Physics, McGill University, Montreal, QC, 
Canada H3A 2T8},
  email={gale@physics.mcgill.ca},
}


\begin{abstract}
We investigate the effects of a charge asymmetry on 
the spectrum of
dileptons radiating from a quark gluon plasma. We demonstrate the existence of 
a new set of processes in this regime. The dilepton production rate from
the corresponding diagrams is
shown to be as important as that obtained from the Born-term quark-antiquark
annihilation.   
\end{abstract}

\date{\today}

\maketitle

\section{Introduction}

The aim of this talk is to show that, when in a medium there 
is a  finite charge density (i.e., a finite chemical
potential), a new
set of lepton pair-producing  processes actually 
arises \footnote{Note that a net charge density in a quark gluon
plasma does not necessarily imply a net baryon density and vice-versa.}. 
We then calculate a new contribution to the
3-loop photon self-energy. 
The various cuts of this self-energy contain higher loop contributions to
the usual processes of $q\bar{q}\rightarrow e^{+}e^{-}$, 
$qg\rightarrow qe^{+}e^{-}$, $q q\rightarrow q q e^{+}e^{-}$, and an
entirely new channel: $gg \rightarrow e^{+}e^{-}$. We calculate the
contribution of this new reaction to the differential production rate 
of back-to-back dileptons. It is finally shown that, within reasonable 
values of parameters,
this process may become larger than the differential rate from the
bare tree level $q\bar{q}\rightarrow e^{+}e^{-}$.     

\begin{figure}
  \resizebox{15pc}{!}{\includegraphics[height=.2\textheight]{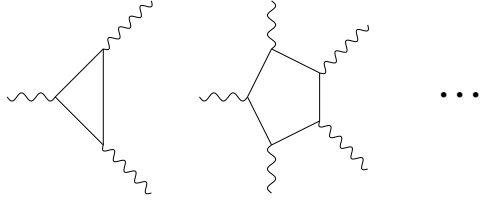}}
  \caption{Diagrams rendered zero by Furry's theorem. See text for details.}
\label{furry}
\end{figure}

At zero
temperature, and at finite temperature and zero charge density (note: henceforth,
a finite density will imply a finite charge density),
diagrams in QED that contain a fermion loop with an odd number of
photon vertices (e.g. Fig. \ref{furry}) are cancelled by an equal and opposite 
contribution coming from the same diagram with fermion lines running in
the opposite direction (Furry's theorem \cite{fur37,itz80,wei95}). This statement can 
also be generalized to QCD for processes with two gluons and an odd 
number of photon vertices.

 A physical perspective is obtained by noting that all these 
diagrams are are encountered in the perturbative evaluation of Green's 
functions with an odd  number of gauge field operators. At zero 
(finite) temperature, in the well defined
 case of QED we observe quantities like 
$\langle 0| A_{\mu1} A_{\mu2} ... A_{\mu2n+1}
 |0\rangle $ ( $ Tr [ \rho(\mu,\beta)  A_{\mu1} A_{\mu2} ... A_{\mu2n+1} ] 
 $ ) under the action of the charge conjugation operator $C$. In QED we know that
 $CA_{\mu}C^{-1} = -A_{\mu} $. In the case of the
 vacuum $|0\rangle $, we note that $C|0\rangle = |0\rangle$, as the vacuum is 
uncharged. As a result 

\vspace{-0.5cm} 

\begin{eqnarray}
 \langle 0| A_{\mu_1} A_{\mu_2} ... A_{\mu_{2n+1}} |0\rangle 
&=&  \langle 0| C^{-1}C A_{\mu_1} C^{-1}C A_{\mu_2} ... A_{\mu_{2n+1}} C^{-1} C |0\rangle
\nonumber \\
= \langle 0| A_{\mu_1} A_{\mu_2} ... A_{\mu_{2n+1}} |0\rangle (-1)^{2n+1} 
&=& -\langle 0| A_{\mu_1} A_{\mu_2} ... A_{\mu_{2n+1}} |0\rangle = 0. 
\end{eqnarray}  

At a temperature $T$, the corresponding quantity to consider is 

\[
\sum_{n} \langle n| A_{\mu_1} A_{\mu_2} ... A_{\mu_{2n+1}} |n\rangle 
e^{-\beta (E_n - \mu Q_n)},
\]

\noindent
where $\beta = 1/T$ and $\mu$ is a chemical potential.
Here, however, 
$C|n\rangle = e^{i\phi}|-n\rangle$, where $|-n\rangle$  is a state in the ensemble with
the same number of antiparticles as there are particles in $|n\rangle$ and vice-versa.
If $\mu = 0$ i.e., the ensemble average displays zero density 
then inserting the operator $C^{-1}C$ as before, we get

\vspace{-0.5cm} 

\begin{eqnarray}
\langle n| A_{\mu_1} A_{\mu_2} ... A_{\mu_{2n+1}} |n\rangle 
e^{-\beta E_n}
&=& - \langle -n| A_{\mu_1} A_{\mu_2} ... A_{\mu_{2n+1}} |-n\rangle 
e^{-\beta E_n}. 
\end{eqnarray}

\noindent The sum over all states will contain the mirror term 
$\langle -n| A_{\mu_1} A_{\mu_2} ... A_{\mu_{2n+1}} |-n\rangle e^{-\beta E_n} $, with
the same thermal weight

\vspace{-0.5cm} 

\begin{eqnarray}
\Rightarrow \sum_{n} \langle n| A_{\mu_1} A_{\mu_2} ... A_{\mu_{2n+1}} |n\rangle 
e^{-\beta E_n } = 0,
\end{eqnarray}

\noindent 
and Furry's theorem still holds. 
However, if
$\mu \neq 0$ 
($\Rightarrow$ unequal number of particles and antiparticles ) 
then

\vspace{-0.5cm} 

\begin{eqnarray}
\langle n| A_{\mu_1} A_{\mu_2} ... A_{\mu_{2n+1}} |n\rangle 
e^{-\beta (E_n - \mu Q_n)}
= - \langle -n| A_{\mu_1} A_{\mu_2} ... A_{\mu_{2n+1}} |-n\rangle 
e^{-\beta (E_n - \mu Q_n)},
\end{eqnarray}

\noindent the mirror term this time is 
$
 \langle -n| A_{\mu_1} A_{\mu_2} ... A_{\mu_{2n+1}} |-n\rangle 
e^{-\beta (E_n + \mu Q_n)}
$
, with a different thermal weight, thus 

\vspace{-0.5cm} 

\begin{eqnarray}
\sum_{n} \langle n| A_{\mu_1} A_{\mu_2} ... A_{\mu_{2n+1}} |n\rangle 
e^{-\beta (E_n - \mu Q_n)} \neq 0,
\end{eqnarray}

\noindent and Furry's theorem will now break down. 
One may say that the medium, being charged, manifestly breaks charge 
conjugation invariance and these
 Green's functions are thus finite, and will lead to the appearance of new processes in
a perturbative expansion. The appearance of processes that can be related to 
 symmetry-breaking in a medium 
  has been noted elsewhere \cite{chi77wel92}.


\begin{figure}[htbp]
  \resizebox{20pc}{!}{\includegraphics[height=.2\textheight]{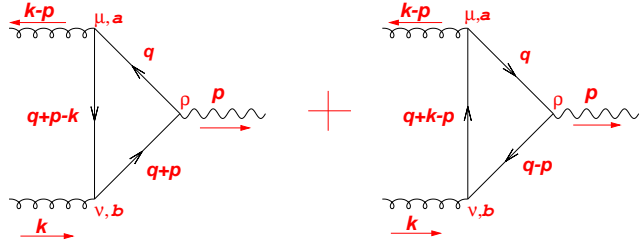}}
    \caption{ The two gluon photon effective vertex as the sum of two diagrams
with quark number running in opposite directions. }
    \label{2vert}
\end{figure}

Let us, now, focus our attention on the diagrams of Fig. \ref{2vert} 
for the case of two gluons and a photon attached to a quark loop 
(the analysis is the
same even for QED i.e., for three photons connected to an electron loop). Such a 
process does not exist at zero temperature, or even at finite 
temperature and zero
 density. At finite density this leads to a new source of dilepton or photon
 production $(gg \rightarrow l^+l^-)$.  
In order
to obtain the full matrix element of a process containing the above as a 
sub-diagram one 
must coherently sum contributions from both diagrams which have fermion number
running in opposite directions. The amplitude for ${\mathcal T}^{\mu \rho \nu}   
( =  T^{\mu \rho \nu} +  T^{\nu \rho \mu} )$  is:

\vspace{-0.5cm} 

\begin{eqnarray}
\!\!\!\!\!T^{\mu \rho \nu} &=& \frac{1}{\beta} \sum_{n= -\infty}^{\infty}
\int^{\infty}_{-\infty} eg^{2} tr[t^{a}t^{b}]
\frac{d^{3}q}{(2\pi)^{3}} Tr[ \gamma^{\mu} \gamma^{\beta} \gamma^{\rho} 
\gamma^{\delta} \gamma^{\nu} \gamma^{\alpha}]  
\frac{(q+p-k)_{\alpha} q_{\beta} 
(q+p)_{\delta} }{(q+p-k)^{2} q^{2} (q+p)^{2}}, \nonumber
\end{eqnarray}    

\begin{eqnarray}
\!\!\!\!\!T^{\nu \rho \mu} &=& \frac{1}{\beta} \sum_{n= -\infty}^{\infty}
 \int^{\infty}_{-\infty} eg^{2} tr[t^{a}t^{b}]
\frac{d^{3}q}{(2\pi)^{3}} Tr[ \gamma^{\nu} \gamma^{\delta} \gamma^{\rho} 
\gamma^{\beta} \gamma^{\mu} \gamma^{\alpha}] 
\frac{(q+k-p)_{\alpha} q_{\beta} 
(q-p)_{\delta} }{(q+k-p)^{2} q^{2} (q-p)^{2}}. \label{2g1p}
\end{eqnarray}

\noindent Again, the extension of Furry's theorem to finite temperature 
does not hold 
at finite density: as, if we set $n \rightarrow -n-1$,
we note that $ q_{0} \rightarrow \!\!\!\!\!\!\!\!\!/ \hspace{0.4mm} -q_{0} $, 
and, as a result, $T^{\mu \rho \nu} (\mu,T) \not= -T^{\nu \rho \mu} (\mu,T)$.
Of course, If we now let the chemical potential go to zero 
($\mu \rightarrow 0$), we note
that for the transformation  $n \rightarrow -n-1$,  we obtain 
$q_{0} \rightarrow  -q_{0}$, and, thus, 
$ T^{\mu \rho \nu} (0,T)  \rightarrow -T^{\nu \rho \mu} (0,T) $. The analysis for
fermion loops with larger number of vertices is essentially the same. 

\vspace{-0.5cm}
\subsection{A Realistic Calculation}

\begin{figure}[htbp]
  \resizebox{14pc}{!}{\includegraphics[height=.2\textheight]{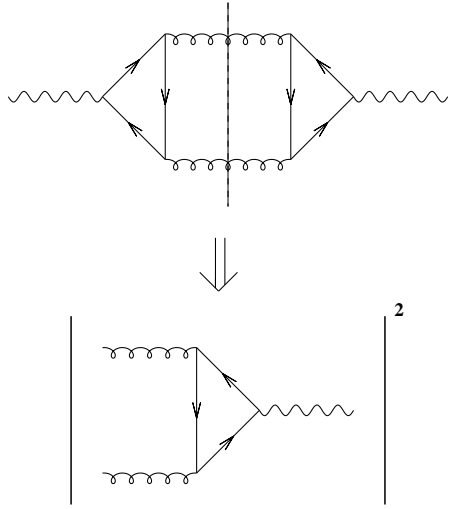}}    
   \caption{The photon self-energy at three loops and the cut that is
evaluated}	
\label{self1E}
\end{figure}

To calculate the contribution made by the diagram of
Fig. \ref{2vert},
to the dilepton spectrum emanating from a quark gluon plasma, we 
calculate the imaginary part of the photon self-energy containing the above diagram
as an effective vertex (Fig \ref{self1E}, see reference \cite{maga2001} for details). 

\begin{equation}
\Pi^{\rho}_{\rho} = \frac{1}{\beta} \sum_{k^{0}} \int \frac{d^{3}k}{(2\pi)^{3}} 
{\mathcal D}_{\eta \mu}(k) {\mathcal T}^{\mu \rho \nu} (k-p,k;p) 
{\mathcal D}_{\nu \zeta}(k-p){\mathcal T}^{\zeta \rho \eta}(k,k-p;-p), 
\end{equation}

\noindent where, the ${\mathcal D}$'s represent bare gluon propagators, and the 
${\mathcal T}$'s represent the effective vertices.
We calculate in the limit of photon three momentum $\vec{p}=0$.
The imaginary part of the considered self-energy contains various cuts. We concentrate,
solely, on the cut that represents the process of 
gluon-gluon to $\mbox{e}^{+}\mbox{e}^{-}$(see Fig.(\ref{self1E})). 
This is a process, which, to our
knowledge, has not been discussed before. The other possible cuts represent
finite density contributions to other known processes of dilepton 
production.

The differential production rate for pairs of massless leptons, with total
energy $E$ and and total momentum $\vec{p}=0$, is given in terms of 
the discontinuity in the
photon self-energy \cite{gal91}, as 

\vspace{-0.5cm}

\begin{equation}
\frac{dW}{dEd^{3}p}(\vec{p}=0) = \frac{\alpha}{12\pi^{3}}\frac{1}{E^{2}}
\frac{1}{1-e^{E/T}}\frac{1}{2\pi i} \mbox{Disc}\Pi_{\rho}^{\rho}(0), \label{rate}
\end{equation}

\noindent 
where $\alpha$ is the electromagnetic coupling constant. The rate of production of a hard
lepton pair with total momentum $\vec{p}=0$, at one-loop order in the photon self-energy 
(i.e., the Born term ), is given for three flavours as  

\begin{equation}
\frac{dW}{dEd^{3}p}(\vec{p}=0) = \frac{5\alpha^{2}}{6\pi^{4}}\tilde{n}(E/2 - \mu)
\tilde{n}(E/2 + \mu) + \frac{\alpha^{2}}{6\pi^{4}}\tilde{n}^2 (E/2)\ .
\label{qqbar}
\end{equation}

\noindent
In the above equation, the first term on the {\it r. h. s.} is the 
contribution from the up and down quark sector; and the second part is the 
contribution from the strange sector. In a realistic plasma, the net charge and 
baryon imbalance is caused by the valence quarks brought in by the incoming charged 
baryon-rich nuclei. The baryon and charge imbalance is, thus, manifested solely in the up and
down quark sector; hence, the chemical potential influences only the distribution function 
of the up and down quarks. The strange and anti-strange quarks are produced in equal 
numbers in the plasma; resulting in a vanishing strange quark chemical potential. 
Thus, dilepton production, from the channel indicated by Fig.(\ref{self1E}), will only
receive contributions from the up and down quark flavours. 

\begin{figure}[htbp]
  \resizebox{12pc}{!}{\includegraphics[height=.2\textheight]{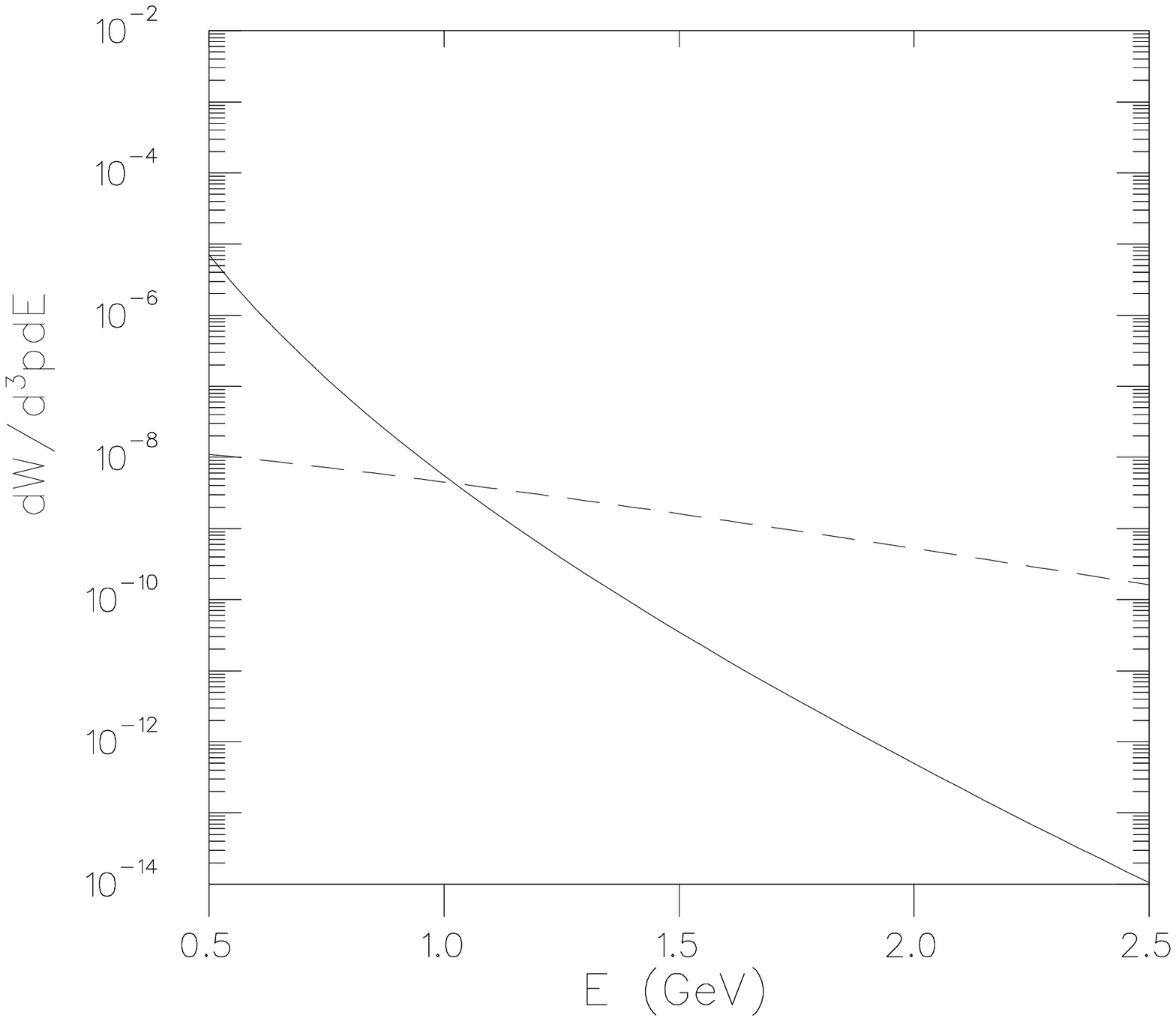}}
   \hspace{1cm}
   \resizebox{12pc}{!}{\includegraphics[height=.2\textheight]{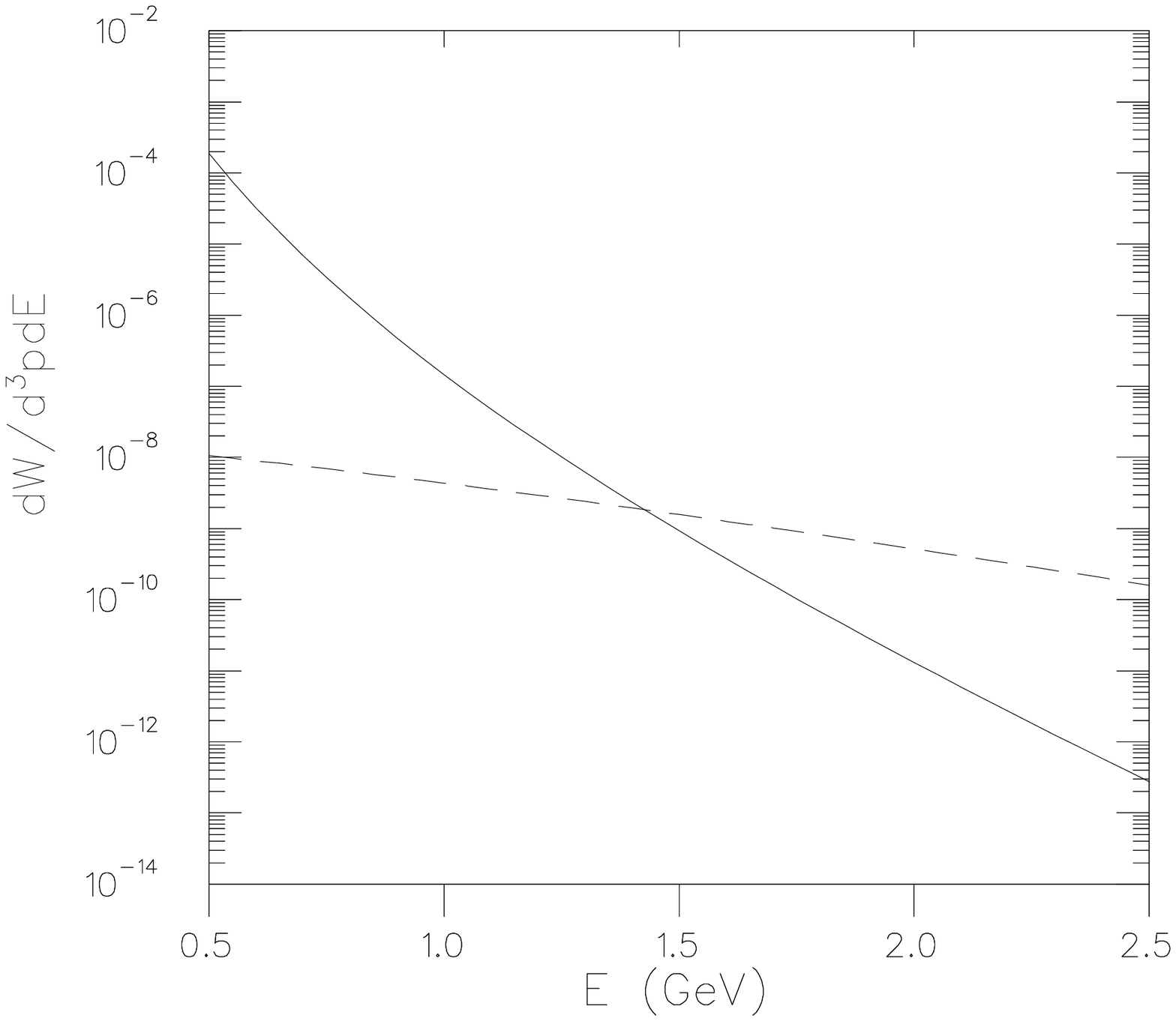}}
    \caption{The differential production rate of back-to-back dileptons
from two processes plotted against dilepton invariant mass. 
The dashed line represents the contribution from Born term 
$q\bar{q}\rightarrow e^{+}e^{-}$. The solid line corresponds to 
the process $gg\rightarrow e^{+}e^{-}$. Temperature is 400 MeV. 
Quark chemical potential, in the first figure, is 0.1T. 
The second figure is the same as the first, but, with
 $\mu=0.5T$  }
    \label{mu}
\end{figure}

\begin{figure}[htbp]
   \resizebox{12pc}{!}{\includegraphics[height=.2\textheight]{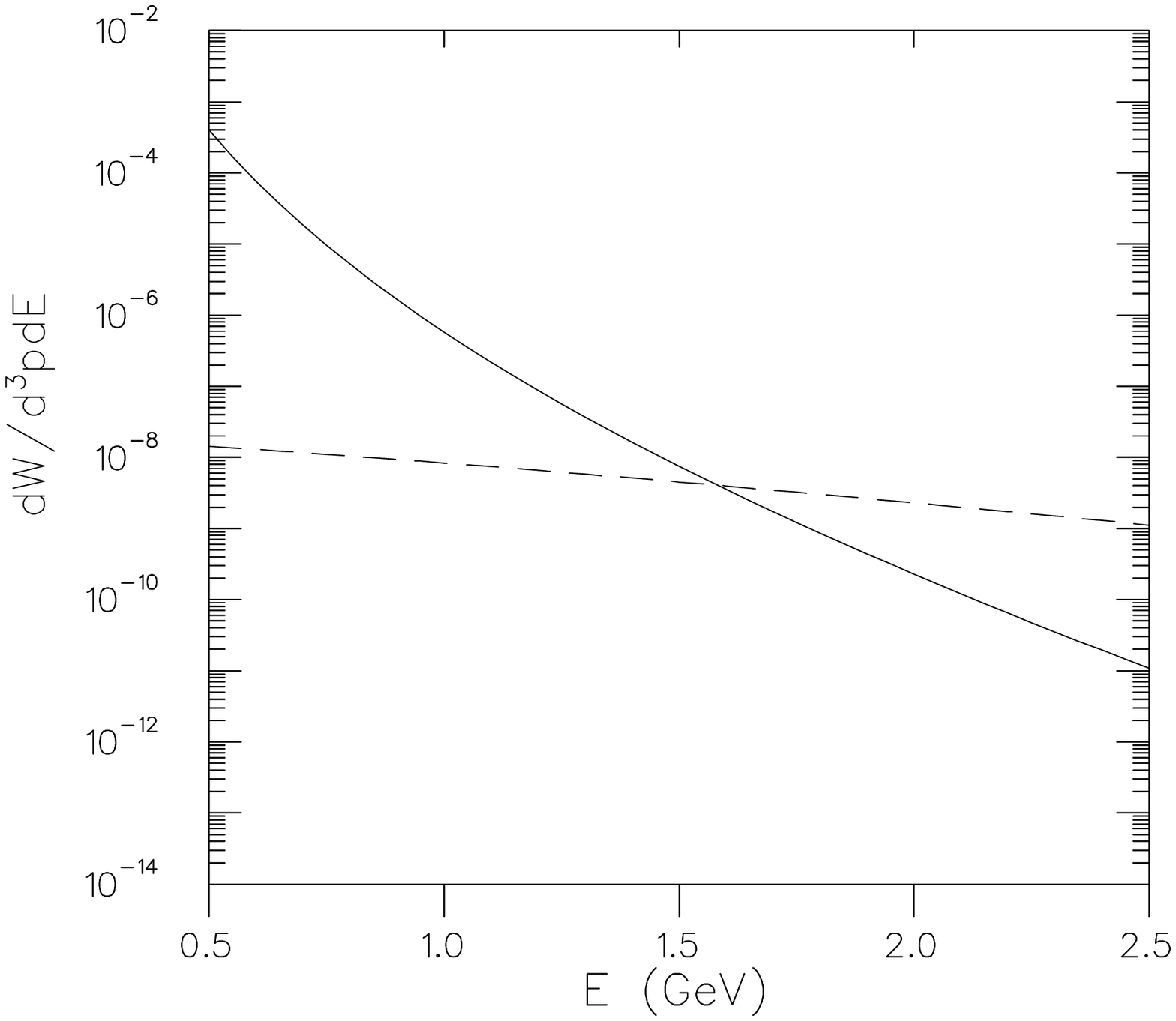}}
   \hspace{1cm}
   \resizebox{12pc}{!}{\includegraphics[height=.2\textheight]{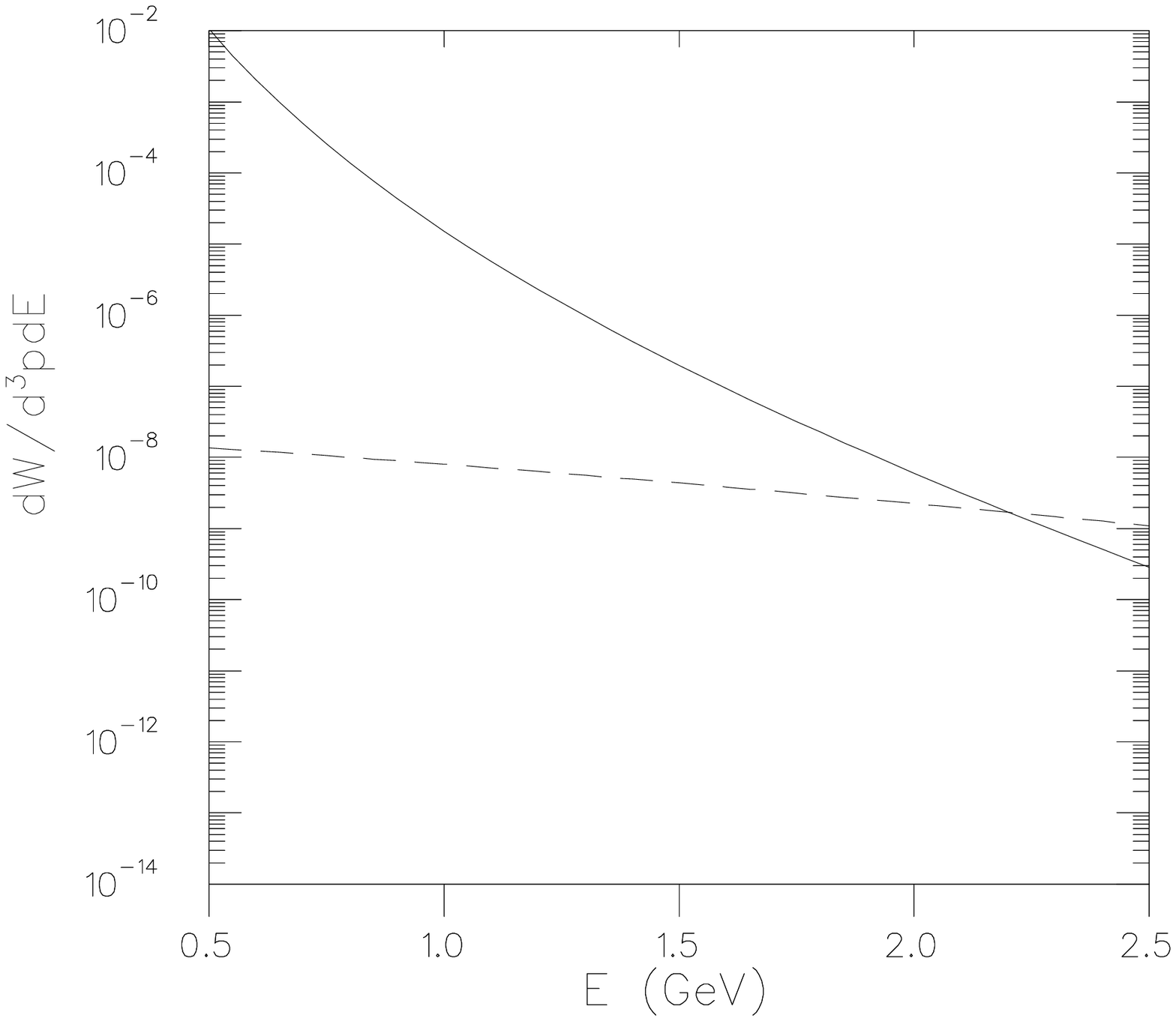}}
   \caption{Same as above but with T=600MeV.}
    \label{nu}
\end{figure}

The initial temperatures of the plasma, formed at RHIC and LHC, 
have been predicted to lie in the range from 300-800 MeV \cite{wan96,rap00}.
 For this calculation 
we use estimates of $T=400\mbox{ MeV}$(Fig. \ref{mu}) and 
$T=600\mbox{ MeV}$(Fig. \ref{nu}). 
To evaluate the effect
 of a finite chemical potential, we perform the calculation  
with two extreme values of
chemical potential $\mu=0.1T$ (left plots) and 
$\mu=0.5T$(right plots)
\cite{gei93}.
The calculation, is performed for three massless flavours of quarks. In this
case, the strong coupling constant is (see \cite{kap00})
 
\begin{equation}
\alpha_{s}(T) = \frac{6\pi}{27 \mbox{ln} ( T/50 \mbox{MeV} )}.
\end{equation}  

\noindent
 The differential rate for the production of dileptons with an invariant mass from 
 0.5 to $2.5\mbox{ GeV}$ is presented. On purpose, we avoid regions where the gluons
 become very soft.
  In the plots, the dashed line is the rate from tree level
 $q\bar{q}$ (Eq. (\ref{qqbar})); the solid line is that from the process 
 $gg\rightarrow e^{+}e^{-}$.  We note that in both cases the gluon-gluon 
process dominates at low energy and dies out at higher energy leaving 
the $q\bar{q}$ process dominant at higher energy. 


\vspace{-0.5cm}
\subsection{High Temperature Limit}

As the reader may have noted, in the above calculation, $\alpha \simeq 0.3,
g \simeq 2$,
 thus, we are not unequivocally in the perturbative regime. 
At asymptotically high temperatures ($T \rightarrow \infty$), 
however, $ g \rightarrow 0$, in this limit one may make 
the Hard Thermal Loop (HTL) approximation \cite{pis2,leb} (note: only the main results
will be quoted here; the details will appear elsewhere \cite{maj01}). At very high
temperature, contributions from loop diagrams are dominated by 
loop momenta $(q)$ of the order of the temperature $T$.
If the momentum flowing in the 
external legs is much smaller (of the order of $gT$, i.e., $q \gg k,p $), then loop
corrections are of the same order in $g$ as the bare diagrams, and, thus, 
must be resummed into the tree amplitudes.

In this limit $(q+p-k)_{\alpha} \simeq q_{\alpha}$, and 
$ E_{q+p} \simeq E_{q} + \vec{p} \cdot \hat{q}  + 
\frac{|\vec{p}|^{2}}{2E_{q}} + 
 \frac{|\vec{p} \cdot \hat{q}|^{2}}{2! \; E_{q}}$. 
On performing the full Matsubara sum and taking the HTL limit for the
numerator, we get  

\vspace{-0.5cm} 

\begin{eqnarray}
{\mathcal T}^{\mu \rho \nu} &=& \int \frac{d^{3}q}{(2\pi)^{3}} 
\frac{\delta^{a b}}{2} e g^{2}
Tr[\gamma^{\mu} \gamma^{\beta} \gamma^{\rho} 
\gamma^{\delta} \gamma^{\nu} \gamma^{\alpha}] 
\frac{\hat{q}_{s_{1},\alpha} \hat{q}_{-s_{2},\beta} 
\hat{q}_{-s_{3},\delta}}
{p^{0} - s_{2}E_{2} + s_{3}E_{3}}  
\nonumber \\
& & \Bigg[ s_{2} \frac{s_{1} 
(\tilde{n}(E_{3}-s_{3}\mu) - \tilde{n}(E_{3}+s_{3}\mu) ) +
s_{3}( \tilde{n}(E_{1}+s_{1}\mu) - \tilde{n}(E_{1}-s_{1}\mu) )}{k^{0} + s_{1}E_{1}
+ s_{3}E_{3}} 
\nonumber \\
&-& s_{3} \frac{s_{1}(\tilde{n}(E_{2}-s_{2}\mu) - 
\tilde{n}(E_{2}+s_{2}\mu)) +
s_{2}(\tilde{n}(E_{1}+s_{1}\mu) - \tilde{n}(E_{1}-s_{1}\mu)) }{k^{0}-p^{0} + 
s_{1}E_{1} + s_{2}E_{2}} \Bigg],  
\end{eqnarray}

\noindent
where, $\hat{q}_{+} = (1,\hat{q}_{1},\hat{q}_{2},\hat{q}_{3})$, and 
$\hat{q}_{-} = (-1,\hat{q}_{1},\hat{q}_{2},\hat{q}_{3})$. Note that if $\mu$ is 
set to zero this amplitude vanishes identically.
The conventional HTL term ( i.e., terms proportional to $(gT)^{2}$ ) 
from  triangle graphs such as these would come from the terms with 
$s_{1} = +,s_{2}= -, s_{3}= -$ and $s_{1} = -,s_{2}= +, s_{3}= +$. 
The term proportional to $(gT)^{2}$ is

\vspace{-0.5cm} 

\begin{eqnarray} 
{\mathcal T}^{\mu \rho \nu}_{(gT)^2} &=& 
\int \frac{d^{3}q}{(2\pi)^{3}} 
\frac{\delta^{a b}}{2} e g^{2} 
Tr[ \gamma ^{\mu} \gamma ^{\beta} \gamma ^{\rho} 
\gamma ^{\delta} \gamma ^{\nu} \gamma ^{\alpha} ] \Bigg \{ 
\frac{\hat{q}_{+,\alpha} \hat{q}_{+,\beta} \hat{q}_{+,\delta}}
{ p^{0} - \vec{p} \cdot \hat{q} } \nonumber \\ 
& & \!\!\!\!\!\!\!\!\!\!\!\!\!\!\!\!\!\!\!\!\!\!
\Bigg[ 
\frac{ 
\frac{\partial \tilde{n}(E_{3}+\mu)}
{\partial E_{3}} (-\vec{k}\cdot\hat{q})
 - \frac{\partial \tilde{n}(E_{3}-\mu)}
{\partial E_{3}}(-\vec{k}\cdot\hat{q} )}
{k^{0} - \vec{k}\cdot\hat{q}} 
+ \frac{ 
\frac{\partial \tilde{n}(E_{2}-\mu)}
{\partial E_{2}} ((\vec{p}-\vec{k})\cdot\hat{q})
- \frac{\partial \tilde{n}(E_{2}+\mu)}
{\partial E_{2}}((\vec{p}-\vec{k}) \cdot \hat{q} )}
{k^{0}-p^{0} + (\vec{p}- \vec{k})\cdot\hat{q}} 
\Bigg] + \nonumber \\ 
& & \!\!\!\!\!\!\!\!\!\!\!\!\!\!\!\!\!\!\!\!\!\!\!\!\!\!\!\!\!
\!\!\!\!\!\!\!\!\!\!\!\!\!\!\!\!\!\!\!\!\!\!\!\!\!\!\!\!\!\!\!
\frac{\hat{q}_{-,\alpha} \hat{q}_{-,\beta} \hat{q}_{-,\delta}}
{p^{0} + \vec{p}\cdot\hat{q}} 
\Bigg[ 
\frac{ 
\frac{\partial \tilde{n}(E_{3}-\mu)}
{\partial E_{3}} (-\vec{k}\cdot\hat{q})
- \frac{\partial \tilde{n}(E_{3}+\mu)}
{\partial E_{3}}(-\vec{k}\cdot\hat{q} )}
{k^{0} + \vec{k}\cdot\hat{q}} 
+ \frac{ 
\frac{\partial \tilde{n}(E_{2}+\mu)}
{\partial E_{2}} ((\vec{p}-\vec{k})\cdot\hat{q})
- \frac{\partial \tilde{n}(E_{2}-\mu)}
{\partial E_{2}}((\vec{p}-\vec{k}) \cdot \hat{q} )}
{ k^{0} - p^{0} - ( \vec{p} - \vec{k} ) \cdot \hat{q} } 
\Bigg]
\Bigg\}. \nonumber
\end{eqnarray}

Note that if we change $\hat{q} \rightarrow -\hat{q}$ in the last two terms 
then $\hat{q}_{+,\alpha} \rightarrow -\hat{q}_{-,\alpha} $ and the whole term becomes
identically zero. Thus there is no term proportional to $(gT)^{2}$ in this diagram. 
The next term is
of order $g^2T$ and is nonzero \cite{maj01}. 
However, as is well known, the HTL approximation of the
$q\bar{q}\gamma$ vertex is of the order of $(gT)^{2}$ \cite{bra90}. 
Thus, in the high temperature approximation, the $gg\gamma$ HTL vertex is suppressed
 as compared to the
$q\bar{q}\gamma$ HTL-resummed vertex. 
As a result the dilepton production rate from resummed 
two-gluon-fusion process 
will also be
suppressed compared to the rate from the resummed Born term, in this
limit.

\begin{figure}[htbp]
  \resizebox{14pc}{!}{\includegraphics[height=.2\textheight,angle=90]{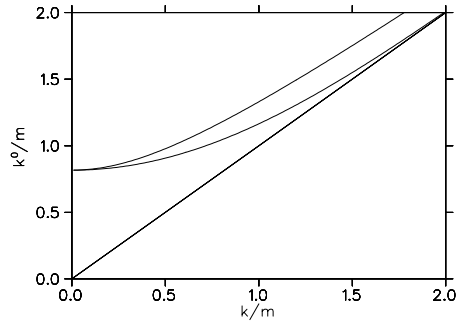}}
    \caption{ HTL resummed dispersion relations of the gluon in a finite temperature,
finite density medium. }
    \label{dsprsn}
\end{figure}

On summing up all the HTL contributions to the self energy of the gluon we get two
dispersion relations for the transverse and longitudinal modes of the gluon. 
A plot of the two modes \cite{maj01} is as shown in Fig. (\ref{dsprsn}). 
The upper branch is the dispersion relation for the transverse quasi-particle
excitation, 
the lower one is the longitudinal excitation. The straight line, corresponding
to free gluons is shown for reference. Note that there is no minima in
either branch ( except at $k=0$ ), unlike in the case of quarks \cite{bra90}. 
Hence, we will
not see any sharp Van Hove peaks \cite{bra90} in the resulting dilepton spectra emanating
from this process. One may thus expect that the high temperature (and, 
perhaps, as a result, the very low invariant mass) dilepton production 
spectra emanating from  
in-medium gluon-gluon fusion will be suppressed compared to that from in-medium
quark anti-quark fusion. However, as we have noted in the previous section the
intermediate invariant mass rate from bare gluon-gluon fusion is comparable and may
be larger than that from bare quark anti-quark fusion. 

The entire calculation above
is at full chemical and thermal equilibrium. In the early part of an
ultrarelativistic heavy ion collision, the gluon number has been predicted to be
much higher than at full chemical equilibrium. In such a scenario, 
the contributions to dilepton
spectrum from processes such as those presented here will probably out-shine those
from other channels.


The authors wish to thank Y. Aghababaie, A. Bourque, S. Das Gupta , F. Gelis,
 S. Jeon, D. Kharzeev,
 C. S. Lam and G. D. Mahlon for 
helpful discussions. A.M. acknowledges the generous support provided to 
him by McGill University through the Alexander McFee fellowship, the Hydro-Quebec
fellowship and the Neil Croll award.
This work was supported in part by the Natural
Sciences and Engineering Research Council of Canada and by { \it le fonds pour la
Formation de Chercheurs et l'Aide \`{a} la Recherche du Qu\'{e}bec. }


\vspace{-0.5cm}

\begin{thebibliography}{99}

\bibitem{maga2001}A. Majumder and C. Gale, Phys. Rev. D, {\bf 63}, 
114008 (2001);and erratum, {\it in press} . 

\bibitem{fur37} W. H. Furry, {\it Phys. Rev.}, {\bf 51}, 125 (1937).

\bibitem{itz80} C. Itzykson, J. B. Zuber, {\it Quantum Field Theory}, McGraw Hill, 
New York, (1980).

\bibitem{wei95} S. Weinberg, {\it The Quantum Theory of Fields}, Vol. 1, Cambridge
University Press, (1995).

\bibitem{chi77wel92} See for example, S. A. Chin, Ann. Phys. 108, 301 (1977); 
H. A. Weldon, Phys. Lett. B, {\bf 274}, 133 (1992); 
G. Wolf, B. Friman and M. Soyeur, Nucl. Phys. {\bf A640}, 129 (1998);
O. Teodorescu, A.K. Dutt-Mazumder and C. Gale, Phys. Rev. C {\bf 63}, 034903 (2001).  
  

\bibitem{gal91} C. Gale, and J. I. Kapusta, Nucl. Phys. {\bf B357}, 65 (1991).
 
\bibitem{wan96} X. N. Wang, Phys. Rep. {\bf280}, 287 (1997).
 
\bibitem{rap00} R. Rapp, hep-ph/0010101.

\bibitem{gei93} K. Geiger, and J. I. Kapusta, Phys. Rev. D. {\bf 47}, 4905 (1993); N.
George, for the PHOBOS collaboration, Proceedings of Quark Matter 2001. 

\bibitem{kap00} J. I. Kapusta, and S. M. H. Wong,  Phys. Rev. C. {\bf62}, 
027901 (2000).   

\bibitem{pis2} E. Braaten, and R. D. Pisarski, Nucl. Phys. {\bf B337} 569 (1990). 

\bibitem{leb} M. Le Bellac, {\it Thermal Field Theory}, Cambridge University
Press, (1996).

\bibitem{hei87} U. Heinz, K. Kajantie, and T. Toimela, Ann. Phys. (N.Y.) {\bf 176} 218 (1987).

\bibitem{bra90} E. Braaten, R. D. Pisarski, and T. C. Yuan, Phys. Rev. Lett. 
{\bf 64} 2242 (1990).

\bibitem{maj01} A. Majumder, and C. Gale, {\it in preparation}. 

\end{thebibliography}
\end{document}